\documentclass[11pt]{article}
\usepackage{lmodern}
\usepackage[T1]{fontenc}
\usepackage[utf8]{inputenc}
\usepackage[margin=1in]{geometry}
\usepackage{amsmath,amssymb,amsthm}
\usepackage{graphicx}
\usepackage{tabularx}
\usepackage[hidelinks]{hyperref}
\usepackage{microtype}
\title{Exact Conditional Distributions of Chi-Square-Family Statistics for Two-Way Contingency Tables, by Cell-Separable Dynamic Programming}
\author{William J. Dwyer, MD, MPH, FAAP\\ Department of Mathematics and Statistics,\\ University of Massachusetts Lowell, Lowell, MA, USA\\ \texttt{wjdwyer@trialdesign.com} \\ ORCID 0009-0004-0855-7222}
\date{}
\begin{document}
\maketitle

\begin{abstract}
Every chi-square-family statistic for a two-way contingency table (Pearson's chi-square, the power-divergence members, the variance-stabilized root statistic) is referred to an approximate null distribution that miscalibrates on sparse or heterogeneous tables, where the exact distribution is a lattice of atoms rather than a continuous curve. The exact reference conditions on both margins (the multivariate Fisher noncentral hypergeometric law) but is usually treated as uncomputable, because the number of tables sharing a margin is astronomical. We show that the exact conditional moments and the exact conditional distribution of any chi-square-family statistic are computable without enumerating tables: the statistic is additive over cells and the margin-conditional law factorizes cell by cell, so a dynamic program walks the table one cell at a time, carrying per state a few moment accumulators, a value-to-probability map, or one complex number, at a cost set by the number of margin states rather than the number of tables. The moment engine handles a five-by-five table at three per cell (about 79 billion tables) in three seconds; the distribution engine returns the exact tail to machine precision, where a three-moment approximation can miss it by 0.28. The engine furnishes Monte-Carlo-free ground truth for benchmarking any approximate reference.
\end{abstract}


\section{Background}

\subsection{The reference is always approximate, and on sparse tables it miscalibrates}

A statistic for testing or measuring association in an R x C contingency table is only as good as the null distribution it is referred to. The chi-square family, $S_{\lambda } = (2 / (\lambda (\lambda +1))) \sum O ((O/E)^{l}ambda - 1)$ (Cressie and Read 1984) and its variance-stabilized relatives, are all referred to a chi-square on $(R-1)(C-1)$ degrees of freedom, or, in more careful work, to a moment-matched chi-square (Patnaik), a saddlepoint, or a nonparametric bootstrap. Each of these is an approximation to a distribution that, on a finite table, is exactly a set of atoms: the statistic takes finitely many values, each with a definite probability, and the "0.95 quantile" of that atomic law is generally not the 0.95 quantile of any continuous curve fitted to it.

On dense, balanced tables the approximation is invisible. On sparse tables (small expected counts) and heterogeneous tables (unequal margins) it is not: the atoms are few and widely spaced, and a continuous reference calibrated to the wrong quantile miscalibrates conservatively for some tables and liberally for others. The companion papers in this series document the consequence for testing (the collapse of the fixed chi-square reference under heterogeneity) and for effect-size intervals (the failure of the noncentral chi-square inversion on sparse tables).

\subsection{The exact distribution is defined, and is usually dismissed as uncomputable}

The exact reference is not mysterious. Condition on both margins, in the tradition of conditional inference (Cornfield 1956). The margin-conditional law of the table is the multivariate Fisher noncentral hypergeometric distribution, $P(O | a, b) proportional to prod_{ij} w_{ij}^{O_{ij}} / prod_{ij} O_{ij}!$, with $w$ the cell odds parameters (all $w = 1$ under independence). This law is exact and finite. The obstacle is that the number of tables sharing a fixed margin is enormous: a 5x5 table at three expected counts per cell has about 79 billion tables on its margin fibre, and a 3x5 table at three per cell has about 2.9 million. Enumeration is out of the question. Exact conditional tests on this fibre have accordingly been implemented by complete enumeration of the tables sharing a margin (Morgan and Blumenstein 1991, for hierarchical log-linear models in multiway tables), definitive, but confined to small tables, and returning a test rather than the conditional distribution or moments of a statistic. For the one-way multinomial, exact goodness-of-fit distributions are nonetheless computed without enumeration, by forming the characteristic function of the additive statistic and inverting it by FFT (Baglivo, Olivier, and Pagano 1992; Keich and Nagarajan 2006) or, recently, by a dynamic program for the Pearson statistic motivated by auditing the continuous approximation (Banic and Elezovic 2025). For the two-way table conditioning on both margins, prior enumeration-avoiding work computes only the null tail of a specific statistic (Pagano and Halvorsen 1981; Mehta and Patel 1983). The exact conditional moments and the full conditional distribution of a general chi-square-family statistic on the two-way margin fibre have not, to our knowledge, been computed without enumeration; that is what we do. For the one-way multinomial the general-family case is by now handled deterministically. Resin (2023), in this journal, computes exact multinomial tests for the whole power-divergence family, but that is the one-way problem, not the two-way both-margins-conditional one. The only existing enumeration-free routes to the two-way both-margins-conditional law are Monte Carlo: sequential importance sampling and Markov-chain sampling (Besag and Clifford 1989; Diaconis and Sturmfels 1998), which return sampled p-values with sampling error, not the exact atomic law or its exact moments. The engine developed here is deterministic and exact, and returns both the exact moments and the full distribution of a general chi-square-family member on that fibre, which no prior deterministic method delivers.

\subsection{The observation that makes it computable}

Two facts change the picture. First, every chi-square-family statistic is \textbf{additive over cells}: $S = sum_{ij} s(O_{ij}, E_{ij})$ for a per-cell contribution $s$. Second, the margin-conditional law \textbf{factorizes cell by cell} as the table is built up subject to the running margin constraints. These are the same two facts, additivity of the statistic and cell-wise factorization of the law, that underlie transform and dynamic-programming methods for exact one-way goodness-of-fit distributions (Baglivo, Olivier, and Pagano 1992; Banic and Elezovic 2025). We apply them to the two-way both-margins-conditional problem: a dynamic program that never materializes a table walks the cells in order, carrying, for each partial state, either a small fixed set of moment accumulators or a map from partial statistic-value to probability. The cost is the number of distinct partial states, not the number of tables, and the two are wildly different: twenty million tables can share a few hundred thousand states. This paper develops that dynamic program, in a moment form and a distribution form, verifies both against enumeration, states what they reach and what they do not, and identifies the one structural result that falls out along the way.

\section{The exact-moment engine}

\subsection{The recursion}

Write $S = sum_{ij} s_{ij}$, $s_{ij} = s(O_{ij}, E_{ij})$. Because the margin-conditional weight and the statistic both factor over cells, the moments of $S$ accumulate by convolution as cells are added. Carrying, per state, the four accumulators

\[
A0 = \sum w, ,\quad A1 = \sum w s, ,\quad A2 = \sum w s^{2}, ,\quad A3 = \sum w s^{3},
\]

and expanding $(s_{new} + s_{partial})^{r}$ on each step gives the exact first three central moments (mean M, variance V, third cumulant K3) of $S$ on the margin fibre. No table is materialized; only the accumulators move.

The naive walk is column by column, and it stalls at 5x5: a column of a 5x5 table at three per cell has 3,876 admissible vectors, each requiring a scatter across the state space, about 152 million inner operations per column, which pure Python and even a naive vectorized scatter cannot afford (the scatter overhead exceeds the arithmetic). The fix is to walk \textbf{one cell at a time} (Figure 1). Then the choice at each step is a single integer, the count in the current cell, and the transition is a shifted slice of a dense array rather than a scatter.

Walking cells appears to need an extra state dimension, how much of the current column is still unplaced, but it does not. Inside column $j$, with $S_{j} = \sum (r)$ the total remaining capacity at the column's start, the count still to place in the column is

\[
col_{left} ,\quad = ,\quad \sum (r) ,\quad - ,\quad (S_{j} - b_{j}),
\]

a function of the current row-capacity vector alone. It carries no information of its own, so the state stays the row-capacity vector and nothing else. Formally: fix a row-major cell order; within each column the count already placed determines $col_{left}$ by the identity above, the last cell of a column is forced by $b_{j}$, and the last column is forced by the row capacities, so every reachable partial table is indexed exactly by the row-capacity vector with no additional axis, and a partial state is admissible if and only if its remaining row capacities can still meet the remaining column totals (the pruning applied at each column boundary). The cost is then $RC x (\max count + 1)$ dense array operations on a lattice of $prod(a_{i} + 1)$ states; the machine-precision agreement with brute-force enumeration in Section 2.2 corroborates the reduction.

\begin{figure}[htbp]\centering
\includegraphics[width=\linewidth]{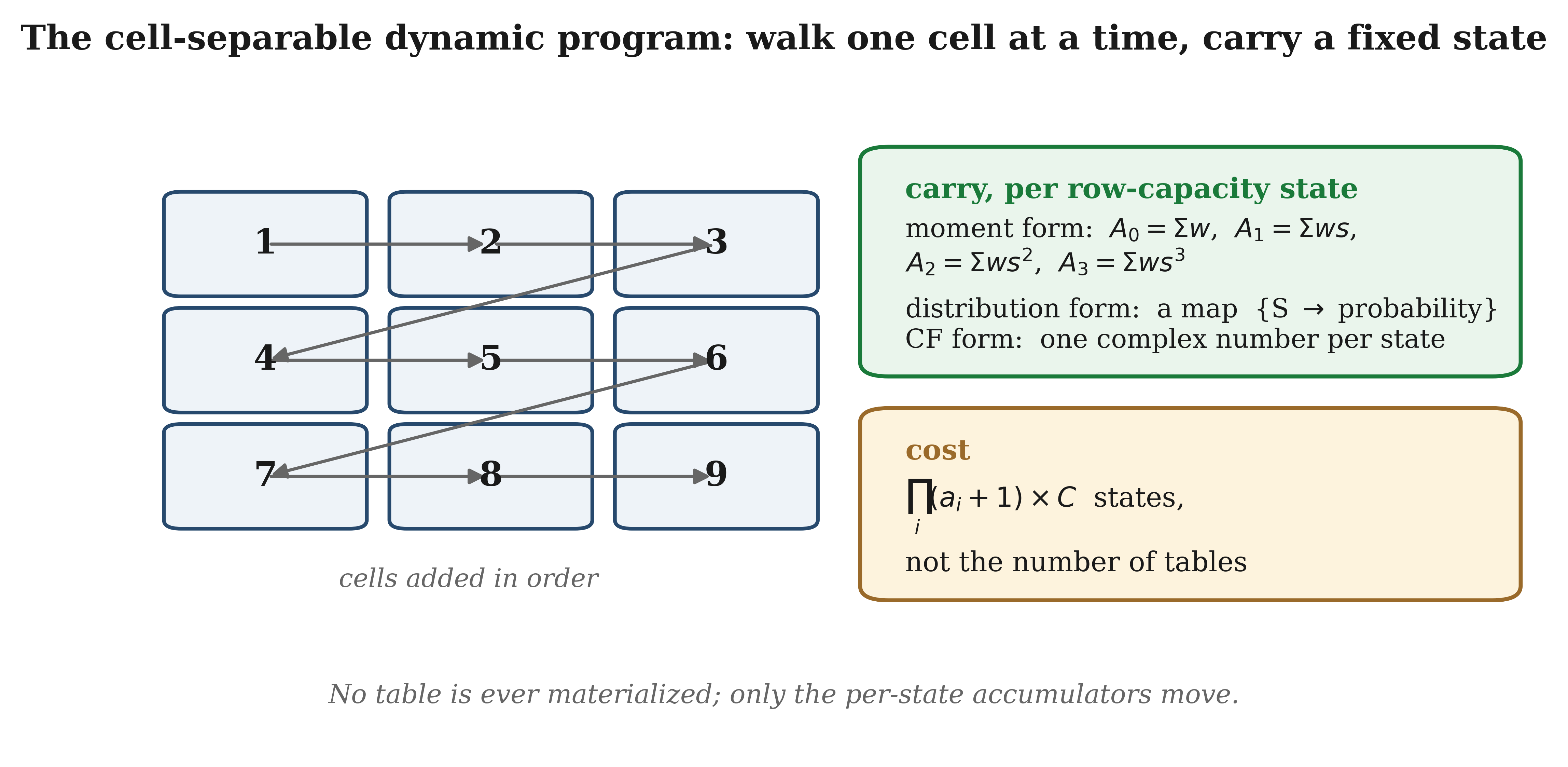}
\caption{The cell-separable dynamic program. The table is built one cell at a time in a fixed order; per row-capacity state the walk carries either four moment accumulators, a map from statistic-value to probability, or one complex number (the characteristic function). No table is ever materialized, and the cost is the number of states, not the number of tables. Generated by rerun/make\_m0h\_figures.py.}
\end{figure}

\subsection{Verification and reach}

Against brute-force enumeration (3x3, N = 18, 406 tables) the moment engine returns the mean, variance, and third cumulant to a maximum relative error of \textbf{8.9e-15} in 5.5 milliseconds. The vectorized engine agrees with the enumeration-verified engine to a worst relative disagreement of \textbf{2.4e-13} across balanced, null, extreme (V = 1.2), heterogeneous, and degenerate (zero-margin) designs.

\begin{table}[htbp]\centering\small
\caption{\textbf{Reach of the moment engine. The cost is states, not tables. Generated by rerun/exact\_moments\_dp.py (vectorized form rerun/exact\_moments\_dp\_fast.py, which reaches the 5x5 timing).}}
\begin{tabular}{lcccc}
\hline
table & N & per cell & tables sharing the margin & time \\
\hline
3x3 & 45 & 5 & 9,316 & 0.08 s \\
3x5 & 45 & 3 & 2,883,031 & 0.11 s \\
4x4 & 48 & 3 & 20,158,151 & 3.3 s \\
5x5 & 75 & 3 & about 79,000,000,000 & 3.0 s (vectorized) \\
\hline
\end{tabular}
\end{table}

The last row is the design at which the standard approximate reference leaks; the exact moments of its 79-billion-table fibre are computed in three seconds. The number of tables is irrelevant (Figure 2). Only $prod(a_{i} + 1) x C$ states, times the counts per cell, matters. The mean and variance of the \emph{cell counts} of this law were already available (Fog 2008); what the recursion adds is the exact moments of the nonlinear additive statistic $S$, which the cell-count moments do not give.

\begin{figure}[htbp]\centering
\includegraphics[width=\linewidth]{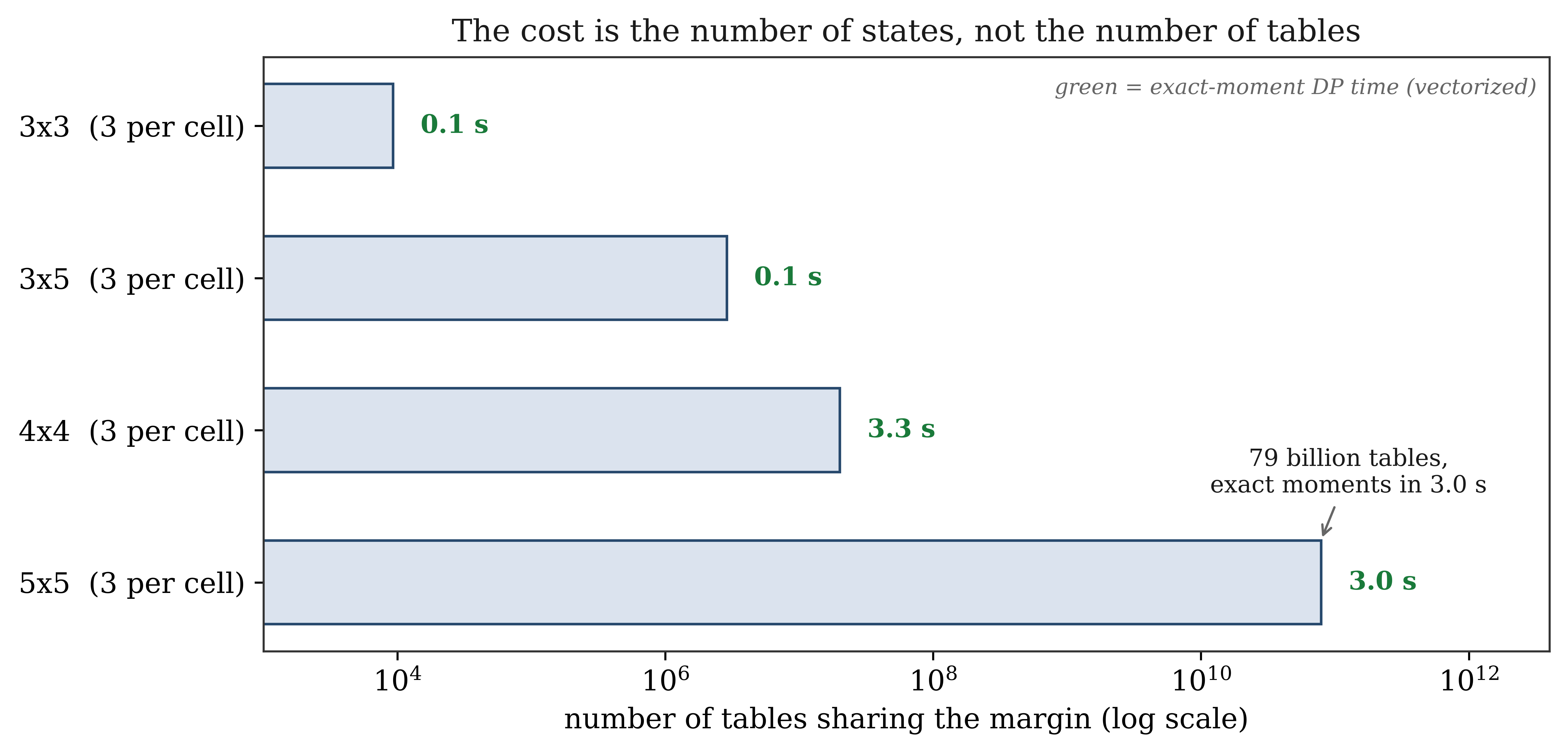}
\caption{The cost is the number of states, not the number of tables. For each design, the bar is the number of tables sharing the margin (log scale) and the green label is the exact-moment dynamic-program time. The 5x5-at-three fibre holds about 79 billion tables; its exact conditional moments are computed in 3.0 seconds. Generated by rerun/make\_m0h\_figures.py.}
\end{figure}

\section{The exact-distribution engine}

\subsection{The moments were not the binding error; the distribution wrapped around them was}

Exact moments do not by themselves give an exact tail. Feeding the exact first three cumulants to a moment-matched (Patnaik) chi-square still produced an interval that covered 0.893 against a nominal 0.95, and on a 2x2 the three-moment chi-square fit missed the true tail by 0.281. The reason is that $S$ is chunky: on a 3x3 at two per cell the margin fibre holds 406 tables, $S$ has 21 atoms, and its true tail at the fitted 0.95 quantile is 0.0390 under the null, not 0.0500; here $S$ is the Anscombe-root statistic the engine carries, and Figure 3 draws the same phenomenon for Pearson $X^{2}$, whose atom count and tail differ. A continuous reference cannot represent an atomic law.

Two sophisticated continuous references were tried and both failed, for the same reason. An exact cumulant-generating-function saddlepoint (Lugannani-Rice) reduced the worst tail error only to 0.035; an exact characteristic function inverted by Gil-Pelaez (in the tradition of continuous quadratic-form inversion; Imhof 1961; Davies 1980; Solomon and Stephens 1977) was worse, up to 0.11, because the characteristic function of a lattice-like variable does not decay and truncating its oscillatory integral rings (a Gibbs effect). That lattice characteristic functions defeat a truncated continuous inversion is well understood in the exact goodness-of-fit literature and is the standard motivation for transform methods that map the statistic to a grid (Baglivo, Olivier, and Pagano 1992; Keich and Nagarajan 2006). Neither continuous reference is wrong because it is imprecise. Both are wrong because they are continuous and the truth is not.

\begin{figure}[htbp]\centering
\includegraphics[width=\linewidth]{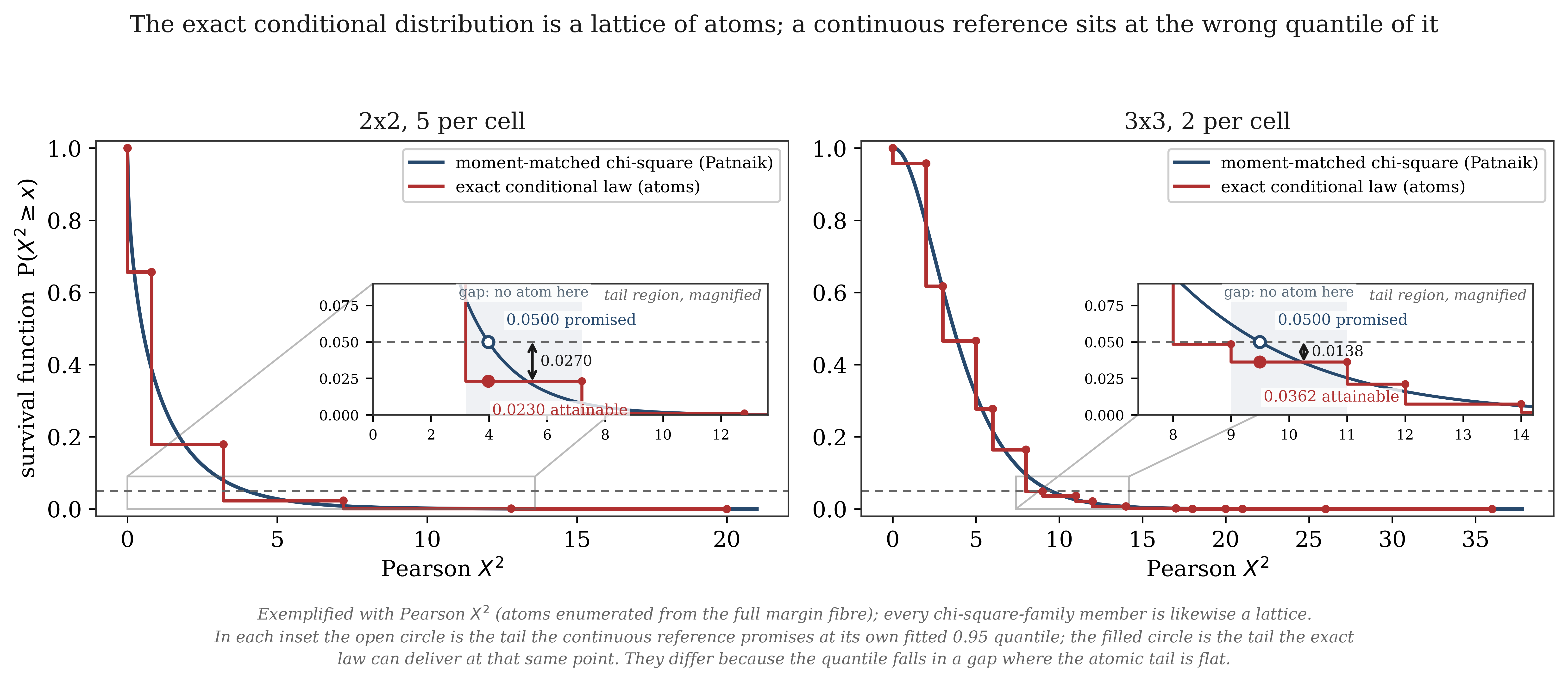}
\caption{Why a continuous reference miscalibrates on a sparse table, shown for Pearson $X^{2}$. The exact conditional law (enumerated from the full margin fibre) is a lattice of atoms; a moment-matched (Patnaik) chi-square sits at the wrong quantile of it, so its nominal 0.95 quantile carries an exact tail different from 0.05. Every chi-square-family member is likewise a lattice. Generated by rerun/make\_m0h\_figures.py.}
\end{figure}

\subsection{Carry the distribution}

Because $S$ is additive over cells and the counts are integers, $S$ takes finitely many values. The same cell-at-a-time walk that carries four moment accumulators can instead carry a map ${S \to probability}$, convolving it as each cell is added and merging coincident atoms. Coincident atoms are identified by rounding each accumulated statistic value to a fixed number of decimals during the walk, coarse enough to absorb floating-point summation-order noise (17.610999999 versus 17.611000001) yet fine enough to keep genuinely distinct atoms apart, which reproduces the exact atomic law on every chi-square-family member tested (Table 2). For the members whose per-cell contribution is a rational function of the integer counts (Pearson, the likelihood ratio, rational-order power-divergence members) the atoms can be keyed in exact rational arithmetic, removing the rounding choice entirely; that is the robust guarantee, and the fixed-decimal key is used only for the variance-stabilized (Anscombe-root) statistic, whose atoms are irrational, with a separation check. The result is the exact conditional law, by construction. The same engine, with the per-cell summand swapped, reproduces the exact conditional law of the Pearson, likelihood-ratio, and Cressie-Read members to machine precision (Table 2), confirming that the recursion handles any additive chi-square-family statistic.

\begin{table}[htbp]\centering\small
\caption{\textbf{The exact-distribution engine against enumeration. Generated by rerun/condexact\_v2\_distengine.py.}}
\begin{tabular}{lcccc}
\hline
design & tables on the fibre & atoms of S & worst tail error & three-moment chi-square error \\
\hline
3x3, 2/cell & 406 & 21 & 4.4e-16 & -0.176 \\
3x3, 3/cell & 1,540 & 56 & 2.2e-16 & -0.117 \\
2x2, 5/cell & 11 & 6 & 1.7e-16 & -0.281 \\
3x3, 2/cell, Pearson $X^{2}$ & 406 & 16 & 3.9e-16 & – \\
3x3, 2/cell, likelihood ratio & 406 & 21 & 4.4e-16 & – \\
3x3, 2/cell, Cressie-Read 2/3 & 406 & 21 & 2.2e-16 & – \\
\hline
\end{tabular}
\end{table}

On a 2x2 the chi-square fit misses the tail by 0.281; the distribution engine misses it by 1.7e-16. One implementation subtlety is worth recording, because an exactness claim that checks only the mean would miss it: rounding the atom keys too finely splits a single atom in two under floating-point summation order (17.610999999 versus 17.611000001), so a tail cut landing between the halves is wrong by up to 0.20 even though total mass and the mean are exact. Rounding coarser than the floating-point noise and finer than the atom spacing fixes it.

\subsection{The characteristic-function variant, for the tables the distribution engine cannot reach}

The distribution engine reaches 4x4 at two per cell and 3x5 at three per cell, but not 5x5 at three per cell, because it carries an extra axis whose length is the number of atoms (2,208 already at 3x5-at-three). The moment engine reaches 5x5 because it carries only accumulators. The characteristic-function route here is the conditional, two-way analogue of the characteristic-function-plus-FFT exact-distribution method of Baglivo, Olivier, and Pagano (1992): to recover the tail at those larger tables without enumerating atoms, the characteristic function $\phi (t) = E[\exp (i t S)]$ is accumulated by the same cell-at-a-time walk carrying \textbf{one complex number per state} rather than a distribution, so it runs on the same row-capacity lattice the moment engine reaches. The exact conditional characteristic function so computed matches direct enumeration to 2.7e-15. Its novelty here is the one-complex-per-state recursion on the both-margins-conditional lattice, and the use of the COS cosine-series inversion (Fang and Oosterlee 2008; for lattice laws, Shen, Fang and Liu 2024) in place of the FFT, which integrates the tail analytically and so avoids the ringing that defeats a truncated Gil-Pelaez integral. It recovers the mid-p tail to about 7e-4 in the upper tail against the exact distribution engine at 3x3; at the minimum atom of a heavy-atom balanced null the truncated cosine window does not resolve the boundary half-atom, so the boundary tail is not delivered to that precision and is not used as a cut point. The characteristic-function route is developed in full, with the interval it induces, in the companion effect-size-interval paper (Dwyer 2026c), whose dedicated coverage sweeps exercise it at 4x4, 5x5 and 6x6; here it is the component of the engine that extends the approximate mid-p tail past the distribution engine's reach.

\section{Exact conditional moments under the alternative, and a leverage identity}

The margin-conditional table is a vector of independent Poisson counts conditioned on the linear margin constraints. Conditioning a Poisson vector on linear constraints projects out the design, so

\[
Cov(O | margins) = W - W X (X' W X)^{-1} X' W, ,\quad W = diag(\mu ),
\]

and the diagonal is $Var(O_{ij} | margins) = mu_{ij} (1 - h_{ij})$, with $h_{ij}$ the diagonal of the hat matrix of the row-plus-column design, that is, the \textbf{row-plus-column leverage}. This is the standard adjusted-residual variance of contingency-table and Poisson/multinomial generalized-linear-model residual theory: Haberman (1973) gives the adjusted residual \texttt{(O\_ij - E\_ij) / sqrt(E\_ij (1 - a\_i/N)(1 - b\_j/N))}, whose denominator variance is exactly \texttt{mu\_ij(1 - h\_ij)}, and the general GLM form $\mu (1 - h)$ with $h$ the hat-matrix leverage is textbook (McCullagh and Nelder 1989). A distinction must be kept precise here. Checked at the null on a balanced 3x3, the leverage gives $1 - h = 0.4444$, which equals the product \texttt{(1 - a\_i/N)(1 - b\_j/N)} (both 4/9), the large-sample adjusted-residual variance factor. The \textbf{exact} conditional variance on the fixed-margin fibre is larger by a finite-population factor: enumeration gives $Var(O_{ij} | margins) = 0.4706 mu_{ij} = mu_{ij}(1 - h_{ij}) N/(N-1)$ at this design. So \texttt{mu\_ij(1 - h\_ij)} is the asymptotic (Haberman 1973) adjusted-residual variance to which the exact conditional variance reduces as $N$ grows, not the exact variance itself, and it is the product form \texttt{(1 - a\_i/N)(1 - b\_j/N) N/(N-1)} of the T\_root reference (its derivation D13, which supersedes the earlier leverage-projection form D9) that carries the finite-population factor and matches enumeration. With that stated, the row-plus-column leverage re-expresses the T\_root variance ratio and ties it to standard generalized-linear-model residual theory (Haberman 1973; McCullagh and Nelder 1989). The identity is textbook; only the interpretive re-identification of the T\_root variance ratio as that leverage is offered here, and it is scaffolding for the section's substantive result below.

The same conditional-moment machinery, carried under an alternative rather than the null, supplies the exact conditional cumulants of the variance-stabilized statistic under a nonzero effect. The centred summand splits as $Z = eps + \beta $, with $\beta $ the noncentrality shift; the even moments involve $\beta ^{2}, \beta ^{4}$ and the \textbf{odd} per-cell moments $m3, m5$ that vanish at the null ($\beta = 0$) and are therefore absent from the null tabulation, plus a noncentral cross term from the trace identity that also vanishes at $\beta = 0$. The derivation is exact and, verified on 3x3, 5x5, 4x6, and 6x6 designs at counts 2 to 10, collapses to the null reference to machine precision at $\beta = 0$. It is the alternative-distribution counterpart of the null reference the T\_root paper uses.

\section{What the engine is worth, and its boundaries}

\subsection{The value: exact ground truth with no Monte Carlo}

Exact computation as a Monte-Carlo-free check on an asymptotic reference is not a new purpose: it is the explicit rationale of the exact one-way goodness-of-fit literature (Baglivo, Olivier, and Pagano 1992; Banic and Elezovic 2025). What is new here is the regime in which that ground truth is now available. Because the engine returns the exact conditional moments and the exact conditional tail for the two-way, both-margins-conditional problem, for the full chi-square family, and under a non-null alternative, the deviation of a Patnaik chi-square, a saddlepoint, a bootstrap, or a closed-form surrogate can be read off to machine precision rather than estimated with Monte Carlo error (Figure 4), in settings the prior one-way null engines did not cover.

\begin{figure}[htbp]\centering
\includegraphics[width=\linewidth]{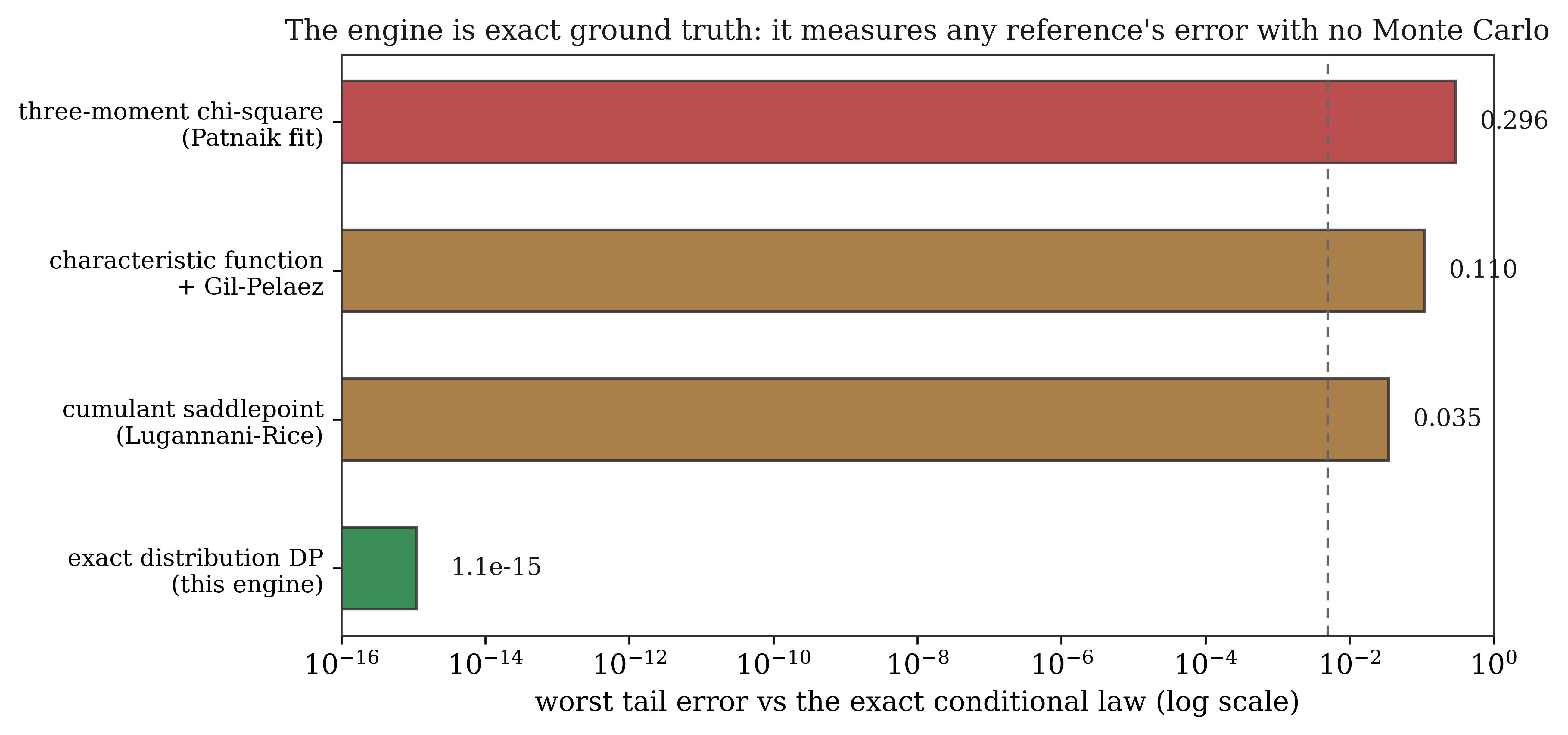}
\caption{The engine as exact ground truth. Worst tail error of each approximate reference against the exact conditional law (log scale): a three-moment (Patnaik) chi-square, a Gil-Pelaez characteristic-function inversion, and a Lugannani-Rice saddlepoint all miscalibrate on a sparse table, while the exact distribution dynamic program agrees with enumeration to machine precision. The engine measures each reference's error with no Monte Carlo. Generated by rerun/make\_m0h\_figures.py.}
\end{figure}

Using it this way we established, exactly, that a closed-form per-cell surrogate for the conditional alternative moments is uselessly wrong on the variance and third cumulant (errors of hundreds to thousands of percent) and does not improve with count, and that even feeding a per-cell binomial model the exact conditional mean and the exact variance ratio still misses the variance by 6.3 percent, because the conditional marginal of a cell is not binomial and the cross-cell dependence under conditioning is stronger than the trace identity allows. The conclusion, reached from measurement rather than argument, is that there is no per-cell surrogate for the conditional alternative: the conditional law has to be used as itself. That is a negative result, and it is exactly the kind of result the exact engine makes it possible to state with certainty.

\subsection{The boundaries, stated plainly}

Three boundaries must be stated rather than blurred. First, \textbf{reach, and what "exact" covers at each tier}: the moment engine returns exact \emph{moments} to 5x5 at three per cell; the distribution engine returns the exact \emph{distribution and tail}, but only to its smaller reach (4x4 at two per cell, 3x5 at three); beyond that the characteristic-function route of Section 3.3 returns a controlled \emph{approximate} mid-p tail (to about 7e-4, with an unresolved boundary half-atom at heavy-atom balanced nulls), not an exact one. So "exact" attaches to the moments wherever the moment engine runs, and to the tail only within the distribution engine's reach; at the frontier the moments are exact and the tail is approximate. Second, \textbf{the residual nuisance}: conditioning on the margins removes the margin nuisance exactly and for free, but a scalar effect size such as $\phi ^{2}$ constrains only one of the $(R-1)(C-1)$ degrees of freedom of the association, so the \textbf{direction} of the departure at a fixed effect size is not fixed. An exact conditional tail computed at the observed direction is a plug-in, not an exact test, and it undercovers (measured coverage 0.931 at a design where the oracle direction gives 0.966 at identical width); the residual direction nuisance has to be handled. In the companion interval paper (Dwyer 2026c) the delivered interval accepts it as an observed-direction (observed-margin) estimand and falls back on a guaranteed projection, with a Berger-Boos supremum as a refinement in the sparse, large-table corner. That companion's dedicated coverage sweeps measure this observed-direction plug-in's coverage across the realized effect size at 4x4, 5x5 and 6x6 (its supplementary coverage tables), confirming undercoverage at small effects and near-nominal coverage by moderate ones, which is why the guaranteed interval there keeps the projection. Third, \textbf{conservatism}: because $S$ is lattice-valued, exactly 0.95 coverage is unattainable, the tail jumps from 0.0390 to 0.0239 with nothing between, so an exact conditional interval is conservative, coverage at least 0.95, in the same way and for the same reason as the Clopper-Pearson interval. That is a guarantee, not a calibration, and any use of the engine to build an interval must say so.

\subsection{A family-wise application: exact simultaneous critical values for interaction-cell residuals}

The engine's exact conditional law is not tied to the chi-square-family sum $S$; it is the joint law of the whole fixed-margin table, so any functional of the residual vector can be referred to it exactly. One functional matters for practice more than the omnibus itself: the \textbf{maximum standardized residual} $M = max_{ij} |r_{ij}|$, whose upper quantile is the critical value for the simultaneous statement "these are the cells that drive the table." A significant omnibus is almost always followed by exactly this localization, and it is where a decomposition reaches the reader's conclusion. The standard practice refers each adjusted residual $r_{ij} = (O_{ij} - E_{ij})/\sqrt{E_{ij}(1 - a_{i}/N)(1 - b_{j}/N)}$ (whose denominator is the leverage variance of Section 4) to a Bonferroni or Sidak-normal cutoff $c = \Phi ^{-1}((1 + (1 - \alpha )^{1/m})/2)$, $m = RC$, which for a 3x3 table at $\alpha = 0.05$ is 2.77. That cutoff assumes the $m$ residuals are independent standard normals. They are neither independent (they share the fixed margins) nor normal (each is a lattice count), and the assumption fails in exactly the sparse, heterogeneous regime where localization is most consequential.

The engine supplies the honest cutoff directly. The exact conditional 1 - alpha quantile of $M$ is read off the same both-margins-conditional law the distribution engine already carries: at 3x3 by enumerating the fixed-margin fibre with its multiple-hypergeometric probabilities and taking the quantile of $M$ atom by atom; beyond 3x3 by a truncation form of the same cell-separable walk, carrying, per state, the mass of partial tables all of whose placed cells satisfy $|r_{ij}| \le c$ (transitions that would violate the bound are zeroed), which accumulates $P(M \le c)$ without enumerating tables, the exact quantile then following from a monotone sweep over $c$. This truncation dynamic program (script \texttt{rerun/m0h\_fwe\_truncation\_dp.py}) is validated to reproduce the enumerated 3x3 quantiles of Table 3 exactly, and it is the same validated recursion that delivers the 4x4 cutoffs of Table 4, where the fixed-margin fibre (of order twenty million tables at three per cell) is far too large to enumerate. Because $M$ is a maximum and not an additive sum, it uses this indicator-truncation recursion rather than the convolution (or the characteristic function) of Section 3. Because the cutoff is the exact conditional quantile, the simultaneous procedure has family-wise error at most $\alpha $ by construction (conservatively so, for the lattice reason of Section 5.2), whereas the asymptotic cutoff is off in both directions and by a wide margin. Table 3 records four sparse heterogeneous 3x3 designs (equal row and column margins shown; every table on the fibre enumerated). At the 25-observation table with margins (20, 3, 2), the Sidak-normal cutoff of 2.77 delivers an actual family-wise error of 8.9 percent, nearly double nominal; the exact conditional cutoff is 3.11 and delivers 2.9 percent. At a 45-observation table the asymptotic procedure's true family-wise error reaches 13.3 percent. The asymptotic cutoff is not merely imprecise: it is anticonservative by roughly a factor of two to three across this corner, and, because on a balanced 15-observation table (5, 5, 5) the same cutoff is instead conservative (exact family-wise error 0.3 percent), there is no fixed correction to the asymptotic bound that repairs it. Only the per-table exact quantile is honest. Figure 5 shows this over a whole grid of designs: panel (a) maps the asymptotic cutoff's achieved family-wise error across sparsity and marginal heterogeneity, and panel (b) confirms that the exact conditional cutoff holds at or below nominal over the same grid.

\begin{table}[htbp]\centering\small
\caption{\textbf{Exact versus asymptotic simultaneous critical values for the maximum adjusted residual (3x3, alpha = 0.05, m = 9; Sidak-normal cutoff 2.77). Generated by rerun/m0h\_fwe\_truncation\_dp.py (enumerated fibre).}}
\begin{tabular}{lccccc}
\hline
margins (row = column) & N & min expected & exact cutoff & family-wise error, Sidak-normal & family-wise error, exact \\
\hline
(5, 5, 5) & 15 & 1.67 & 2.71 & 0.003 & <= 0.05 \\
(14, 3, 3) & 20 & 0.45 & 2.87 & 0.060 & 0.006 \\
(20, 3, 2) & 25 & 0.16 & 3.11 & 0.089 & 0.029 \\
(22, 3, 3) & 28 & 0.32 & 3.32 & 0.094 & 0.022 \\
(40, 3, 2) & 45 & 0.09 & 3.20 & 0.133 & 0.029 \\
\hline
\end{tabular}
\end{table}

The same behaviour, and the same fix, extend past the reach of enumeration. Table 4 reports four 4x4 designs ($m = 16$ cells, Sidak-normal cutoff 2.95), whose fixed-margin fibres are too large to enumerate and whose exact cutoffs are therefore computed by the truncation dynamic program above, the recursion validated against the enumerated 3x3 numbers of Table 3. The pattern is identical: on the balanced 24-observation table (6, 6, 6, 6) the Sidak-normal cutoff is mildly conservative (exact family-wise error 1.3 percent), while on the sparse, heterogeneous 40-observation table (30, 5, 3, 2) it delivers a true family-wise error of 22.3 percent (more than four times nominal) against an exact conditional cutoff of 3.43 that holds at 4.8 percent. As at 3x3, the asymptotic bound is conservative in one corner and grossly anticonservative in another, so no fixed inflation repairs it; only the per-table exact quantile, now reached without enumeration, is honest.

\begin{table}[htbp]\centering\small
\caption{\textbf{Exact versus asymptotic simultaneous critical values for the maximum adjusted residual (4x4, alpha = 0.05, m = 16; Sidak-normal cutoff 2.95), cutoffs computed by the truncation dynamic program validated in Table 3. Generated by rerun/m0h\_fwe\_truncation\_dp.py (truncation dynamic program).}}
\begin{tabular}{lcccc}
\hline
margins (row = column) & N & exact cutoff & family-wise error, Sidak-normal & family-wise error, exact \\
\hline
(6, 6, 6, 6) & 24 & 2.72 & 0.013 & 0.013 \\
(20, 4, 3, 3) & 30 & 3.45 & 0.101 & 0.009 \\
(30, 5, 3, 2) & 40 & 3.43 & 0.223 & 0.048 \\
(40, 4, 3, 3) & 50 & 3.86 & 0.118 & 0.032 \\
\hline
\end{tabular}
\end{table}

\begin{figure}[htbp]\centering
\includegraphics[width=\linewidth]{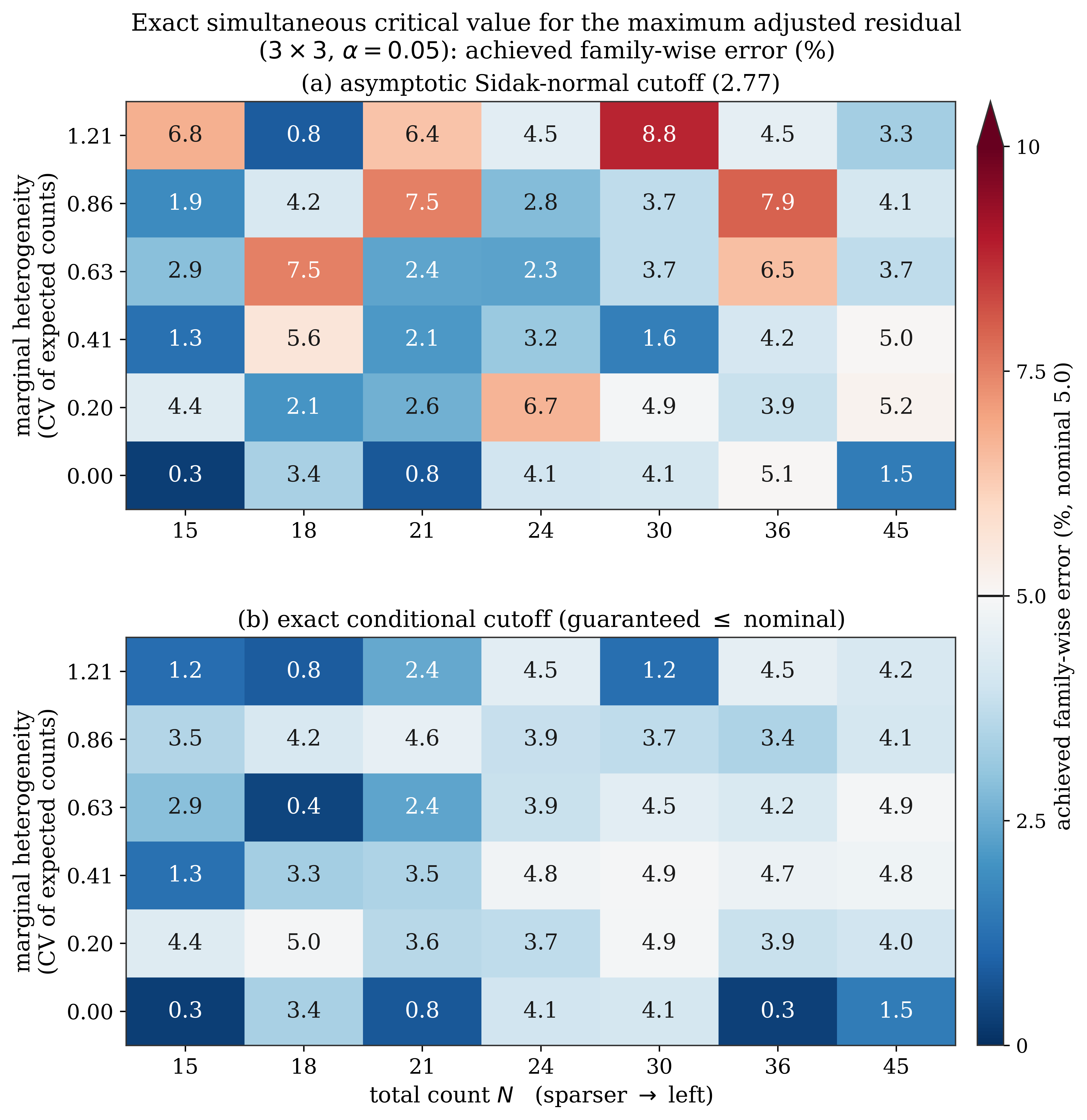}
\caption{The family-wise guarantee. Achieved family-wise error of the simultaneous procedure for the maximum adjusted residual over a grid of 3x3 designs (horizontal axis, total count N, a sparsity proxy; vertical axis, marginal heterogeneity measured by the coefficient of variation of the expected counts), every value computed exactly by enumerating the fixed-margin fibre with its multiple-hypergeometric probabilities. Panel (a): the asymptotic Sidak-normal cutoff reaches 8.8 percent in the sparse, heterogeneous corner and falls to 0.3 percent at the balanced table, so it is anticonservative and conservative in different places and no single fixed correction repairs it. Panel (b): the exact conditional cutoff is at most the nominal 5 percent everywhere, by construction (grid maximum 4.95 percent, below nominal for the lattice reason of Section 5.2). Generated by rerun/make\_m0h\_fig5\_fwe.py.}
\end{figure}

The grid in Figure 5 makes the point that a single worked table cannot: the asymptotic bound's error is not a smooth function that a fixed inflation factor could absorb, but scatters between 0.3 and 8.8 percent as the count and the margins move (panel a), precisely because the residual vector's joint law is the discrete fixed-margin law and not a fixed Gaussian. The exact conditional cutoff tracks that law table by table and so stays at or below nominal across the whole plane. This is the exact-conditional route to which a companion reporting standard for interaction decomposition (Dwyer, in preparation) points its \emph{simultaneous} localization claim: one-at-a-time cell reporting needs only per-component marginal calibration, but a joint "which set of cells" statement needs the components mutually calibrated, and under sparsity that is delivered honestly only by the exact conditional joint law, not by a variance-stabilized residual referred to an asymptotic bound. A separate line of work asked whether the variance-stabilized (Anscombe) residual would make the simultaneous cutoff portable near the Sidak-normal value; a coverage study established that it does not: the stabilized max residual keeps a heavier upper tail because a near-empty cell's stabilized residual does, so its exact cutoff drifts further from the asymptotic bound, not closer. The variance-stabilization advantage is a second-moment, whole-partition property; the maximum over cells is a single-cell tail property, and the two do not coincide. The honest simultaneous procedure is therefore an application of this engine, not of a stabilized residual, and it lives entirely inside the boundaries of Section 5.2: an exact, conservative, guaranteed family-wise statement, with no Monte Carlo and no asymptotic cutoff.

\subsection{The interval companion, and a boundary: exact simultaneous odds ratios for cells}

Section 5.3 answers \emph{which} cells drive the table. The paired question is \emph{how strongly}, which asks for an interval, and it turns out to mark a boundary of this engine: a place where, unlike Section 5.3, the exact machinery earns little, which is worth stating plainly. The natural per-cell effect size is the odds ratio $psi_{ij}$ of the cell-versus-rest 2x2 collapse, a fixed, margin-invariant parameter, for which the cell count given both margins is Fisher noncentral hypergeometric. Inverting that exact conditional test gives the Cornfield exact interval for $psi_{ij}$, and a Bonferroni allocation across the $m$ cells gives a simultaneous band that covers the whole odds-ratio vector with probability at least $1 - \alpha $ by construction. In simulation from tables with genuine association (2000 replicates per design; Monte-Carlo standard error about 0.003 at these levels) this exact band holds family-wise coverage, conservatively, at 0.985 to 0.995. The point of the boundary is that the standard asymptotic competitor is \emph{also} adequate: the Haldane-corrected Wald log-odds-ratio band with a Bonferroni-normal critical value holds family-wise coverage at or above nominal across the same designs (0.949 to 0.975, checked down to fifteen observations with half the intervals one-sided). So, in sharp contrast to the \emph{test} of Section 5.3 (where the asymptotic bound is off in both directions and the exact conditional cutoff is essential), there is no coverage gap to close on the interval side; both the exact and the asymptotic simultaneous odds-ratio bands are valid.

Two negatives make the boundary precise. First, the exact conditional \emph{joint} law does not sharpen the multiplicity below Bonferroni: calibrating a common per-cell level from the conditional minimum-p statistic lands \emph{below} \texttt{alpha/m} in the sparse discrete regime (0.87 of it at 3x3, as little as 0.14 at 4x4), so the joint-law band is 2 to 22 percent wider, not narrower, at equal coverage. Second, the exact per-cell intervals are wide and often one-sided on sparse cells (a third to a half of them), an intrinsic property of exact odds-ratio information, not a defect. The reading is that the engine's joint machinery pays for the family-wise cutoff of a shared statistic (Section 5.3) and earns nothing for a vector of separately-exact per-cell intervals, where a Haldane-Wald band or a Bonferroni union bound already suffices (Section 5.4). This is the honest edge of the method, and it is more useful to locate it than to manufacture a gap: the full development, with the odds-ratio worked example on the oesophageal-cancer case-control data of Breslow and Day (1980), is a companion note (Dwyer 2026f).

\section{Relation to the rest of the series}

This engine is the computational substrate of the series, not a standalone application. The T\_root testing papers use its exact conditional moments as the reference their closed-form three-moment match approximates, and Section 4's leverage identity and alternative-moment derivation are the exact objects that closed form targets. The effect-size-interval paper (Dwyer 2026c) uses the exact conditional distribution of Section 3, and the characteristic-function extension of Section 3.3, to build the exact conditional confidence interval for Cramer's V, with the residual direction nuisance of Section 5.2 accepted as the observed-direction estimand and bounded by a guaranteed projection (a Berger-Boos supremum refining it in the sparse corner). Presented on its own, the contribution is the algorithm and its verified reach: that the exact conditional moments and the exact conditional distribution of a chi-square-family statistic are computable at a cost set by the margins rather than the table count, that the computation is exact to machine precision, and that it furnishes Monte-Carlo-free ground truth for the two-way both-margins-conditional problem, against which every approximate reference in this literature can be measured. A further corollary serves the series' component-partition work. The moment recursion applies to any statistic additive over cells, and a single interaction component of the Lancaster decomposition, $g = sum_{ij} (F_{ir} H_{js} / \sqrt{E_{ij}}) (O_{ij} - E_{ij})$, is a \emph{linear} additive statistic; carrying the same per-row-capacity accumulators one order further returns the exact both-margins-conditional mean, variance, and third and fourth cumulants of $g$ with no enumeration, at the same per-state cost. In particular the exact conditional fourth cumulant of a component, equivalently $Var(g^{2}) - 2$, is obtained for tables far beyond enumeration. The only change to the algorithm is that the power-divergence per-cell contribution is replaced by the linear one and the accumulators run from three moments to four; the result is verified against brute-force enumeration on small tables to machine precision, and is the exact reference a companion reporting standard (Dwyer, in preparation) uses to set a per-component cutoff without simulation.

\section*{Supplementary materials and reproducibility}

All code is deposited and is also provided as an installable package, $condexact$, which exposes the engines through a documented public API ($conditional_{moments}$, $conditional_{distribution}$, $conditional_{tail}$, \texttt{cramer\_v\_interval}) and ships a test suite that verifies the exactness claims against brute-force enumeration. The package vendors the validated engines: the moment engine (\texttt{exact\_moments\_dp.py}) and its vectorized form (\texttt{exact\_moments\_dp\_fast.py}), the distribution engine (\texttt{dist\_engine.py}), the characteristic-function engine and its COS inversion (\texttt{cf\_tail.py}, \texttt{cos\_tail.py}), and the alternative-moment derivation (\texttt{anscombe\_alt.py}, \texttt{v\_root\_alt.py}). Every number regenerates from the deposited code and is checked against brute-force enumeration where enumeration is feasible. The derivations are consolidated in the companion derivations appendix (DH1 to DH8), included with the deposit, which draws together the working notes behind each engine. The reproducibility package is archived at Zenodo under the concept DOI 10.5281/zenodo.21831680 (this release is version 1.0.0, 10.5281/zenodo.21831681).

\textbf{Disclosure statement.} The author develops and hosts the open-source software and associated web domains (the trialdesign.com applications) that implement the methods described, including the condexact package; no other competing interests are declared.

\textbf{Funding.} None.

\textbf{Data availability statement.} The data and code that support the findings of this study are openly available in Zenodo at https://doi.org/10.5281/zenodo.21831680. All results are computed by deterministic code (the $condexact$ package) archived in that record; the contingency tables analyzed are simulated and enumerated by the deposited scripts, so no third-party or human-subjects data were used. Every numerical result regenerates from the deposited code and is verified against brute-force enumeration where enumeration is feasible.

\textbf{Use of generative AI.} In preparing this manuscript and its accompanying software, the author used the generative-AI assistant Claude (Anthropic; Claude Opus 4, accessed via the Claude desktop application in 2026). It was used for copyediting and language refinement, drafting and revising prose, formatting, and coding assistance for the deposited package; it was not used to generate research results, data, or references. All AI-assisted output was reviewed, verified, and edited by the author, who takes full responsibility for the accuracy and integrity of the content.

\textbf{Provenance of results.} Every numerical result is produced by deterministic, human-reviewed code and is verified against brute-force enumeration where enumeration is feasible; the exactness claims are machine-precision agreements with that enumeration, not Monte Carlo estimates.

\section*{References}

\noindent Baglivo, J., Olivier, D., \& Pagano, M. (1992). Methods for exact goodness-of-fit tests. Journal of the American Statistical Association, 87(418), 464-469.\par\smallskip

\noindent Banic, N., \& Elezovic, N. (2025). Zero-disparity distribution synthesis: fast exact calculation of chi-squared statistic distribution for discrete uniform histograms. Preprint, arXiv:2506.23416.\par\smallskip

\noindent Besag, J., \& Clifford, P. (1989). Generalized Monte Carlo significance tests. Biometrika, 76(4), 633-642.\par\smallskip

\noindent Breslow, N. E., \& Day, N. E. (1980). Statistical Methods in Cancer Research, Volume 1: The Analysis of Case-Control Studies. IARC Scientific Publications No. 32, Lyon.\par\smallskip

\noindent Cornfield, J. (1956). A statistical problem arising from retrospective studies. Proceedings of the Third Berkeley Symposium on Mathematical Statistics and Probability, 4, 135-148.\par\smallskip

\noindent Cressie, N., \& Read, T. R. C. (1984). Multinomial goodness-of-fit tests. Journal of the Royal Statistical Society, Series B, 46(3), 440-464.\par\smallskip

\noindent Davies, R. B. (1980). The distribution of a linear combination of chi-squared random variables. Applied Statistics, 29(3), 323-333.\par\smallskip

\noindent Diaconis, P., \& Sturmfels, B. (1998). Algebraic algorithms for sampling from conditional distributions. The Annals of Statistics, 26(1), 363-397.\par\smallskip

\noindent Dwyer, W. J. (2026). T\_root: a comprehensive closed-form statistic for independence in sparse and heterogeneous contingency tables. Companion manuscript, submitted for publication.\par\smallskip

\noindent Dwyer, W. J. (2026c). Exact conditional confidence intervals for Cramer's V: near-nominal and tight for contingency-table effect sizes. Companion manuscript, submitted for publication.\par\smallskip

\noindent Dwyer, W. J. (2026f). Exact simultaneous confidence intervals for the cell odds ratios of a sparse contingency table. Companion note, this series.\par\smallskip

\noindent Dwyer, W. J. (in preparation). A reporting standard for interaction decomposition of sparse contingency tables. Manuscript in preparation.\par\smallskip

\noindent Fang, F., \& Oosterlee, C. W. (2008). A novel pricing method for European options based on Fourier-cosine series expansions. SIAM Journal on Scientific Computing, 31(2), 826-848.\par\smallskip

\noindent Fog, A. (2008). Sampling methods for Wallenius' and Fisher's noncentral hypergeometric distributions. Communications in Statistics - Simulation and Computation, 37(2), 241-257.\par\smallskip

\noindent Haberman, S. J. (1973). The analysis of residuals in cross-classified tables. Biometrics, 29(1), 205-220.\par\smallskip

\noindent Imhof, J. P. (1961). Computing the distribution of quadratic forms in normal variables. Biometrika, 48(3-4), 419-426.\par\smallskip

\noindent Keich, U., \& Nagarajan, N. (2006). A fast and numerically robust method for exact multinomial goodness-of-fit test. Journal of Computational and Graphical Statistics, 15(4), 779-802.\par\smallskip

\noindent Lugannani, R., \& Rice, S. (1980). Saddle point approximation for the distribution of the sum of independent random variables. Advances in Applied Probability, 12(2), 475-490.\par\smallskip

\noindent McCullagh, P., \& Nelder, J. A. (1989). Generalized Linear Models (2nd ed.). Chapman and Hall.\par\smallskip

\noindent Mehta, C. R., \& Patel, N. R. (1983). A network algorithm for performing Fisher's exact test in r x c contingency tables. Journal of the American Statistical Association, 78(382), 427-434.\par\smallskip

\noindent Morgan, W. M., \& Blumenstein, B. A. (1991). Exact conditional tests for hierarchical models in multidimensional contingency tables. Journal of the Royal Statistical Society, Series C (Applied Statistics), 40(3), 435-442.\par\smallskip

\noindent Pagano, M., \& Halvorsen, K. T. (1981). An algorithm for finding the exact significance levels of r x c contingency tables. Journal of the American Statistical Association, 76(376), 931-934.\par\smallskip

\noindent Patnaik, P. B. (1949). The non-central chi-square and F-distributions and their applications. Biometrika, 36(1-2), 202-232.\par\smallskip

\noindent Resin, J. (2023). A simple algorithm for exact multinomial tests. Journal of Computational and Graphical Statistics, 32(2), 539-550.\par\smallskip

\noindent Shan, G., \& Gerstenberger, S. (2017). Fisher's exact approach for post hoc analysis of a chi-squared test. PLOS ONE, 12(12), e0188709.\par\smallskip

\noindent Shen, Y., Fang, G., \& Liu, Y. (2024). Fourier-cosine method for evaluating distributions of discrete random variables. arXiv:2410.04487.\par\smallskip

\noindent Solomon, H., \& Stephens, M. A. (1977). Distribution of a sum of weighted chi-square variables. Journal of the American Statistical Association, 72(360), 881-885.\par\smallskip

\noindent Westfall, P. H., \& Young, S. S. (1993). Resampling-Based Multiple Testing: Examples and Methods for p-Value Adjustment. Wiley.\par\smallskip

\noindent Zhang, Q. (2024). On the asymptotic distributions of some test statistics for two-way contingency tables. arXiv:2409.14255.\par\smallskip

\end{document}